\documentclass[aps,prb,groupedaddress,showpacs,twocolumn,superscriptaddress]{revtex4}
\usepackage{graphicx}
\usepackage{amsmath}
\usepackage{amssymb}
\usepackage{bbm}
\usepackage{xspace}
\usepackage{color}
\usepackage{footnote}
\usepackage{comment}
\usepackage{hyperref}

 \usepackage[utf8]{inputenc}

\definecolor{orange}{RGB}{252,77,6}
\definecolor{brown}{RGB}{200,127,50}
\definecolor{blue}{RGB}{00,000,100}
\definecolor{green1}{RGB}{00,100,00}
\definecolor{green2}{RGB}{00,150,00}
\definecolor{green3}{RGB}{00,200,00}
\definecolor{green4}{RGB}{00,250,00}
\definecolor{grey}{RGB}{100,100,100}

\newcommand{\fig}[1]{Fig.\thinspace{}\ref{#1}}

\newcommand{\eq}[1]{Eq.\thinspace{}(\ref{#1})}

\newcommand{\eqs}[1]{Eqs.\thinspace{}(\ref{#1})}

\newcommand{\se}{Sec.\@\xspace}

\newcommand{\app}{App.\@\xspace}

\newcommand{\tcite}[1]{Ref.~\onlinecite{#1}}
\newcommand{\tcites}[1]{Refs.~\onlinecite{#1}}

\newcommand{\und}{\underline}
\newcommand{\iim}{\Im{}\,}

\newcommand{\up}{\ensuremath{\uparrow}}
\newcommand{\dw}{\ensuremath{\downarrow}}

\def\Braket#1{\mathinner{\left<{#1}\right>}}

\newcommand{\nag}{{\phantom{\dagger}}}

\newcommand{\DT}{$\Delta{T}$ }
\newcommand{\DTns}{$\Delta{T}$}
\newcommand{\dDT}{$d(\Delta{T})$ }
\newcommand{\dDTns}{$d(\Delta{T})$}
\newcommand{\dT}{\Delta{T}}
\newcommand{\djdT}{$\partial{j}/\partial(\dT)$ }
\newcommand{\djdV}{$\partial{j}/\partial\phi$ }
\newcommand{\djdTns}{$\partial{j}/\partial(\dT)$}
\newcommand{\djdVns}{$\partial{j}/\partial\phi$}

\newcommand{\footnoteremember}[2]{
\footnote{#2}
\newcounter{#1}
\setcounter{#1}{\value{footnote}}
}
\newcommand{\footnoterecall}[1]{
\cite{endnote\arabic{#1}} 
}

\begin{document}

\title{Thermoelectric response of a correlated impurity in the nonequilibrium Kondo regime}

\author{Antonius Dorda}
\email[]{dorda@tugraz.at}
\affiliation{Institute of Theoretical and Computational Physics, Graz University of Technology, 8010 Graz, Austria}
\author{Martin Ganahl}
\affiliation{Perimeter Institute for Theoretical Physics, Waterloo, Ontario, N2L 2Y5, Canada}
\author{Sabine Andergassen}
\affiliation{Institut f\"ur Theoretische Physik and Center for Quantum Science,\\Universit\"at T\"ubingen, Auf der Morgenstelle 14, 72076 T\"ubingen, Germany}
\author{Wolfgang von der Linden}
\affiliation{Institute of Theoretical and Computational Physics, Graz University of Technology, 8010 Graz, Austria}
\author{Enrico Arrigoni}
\email[]{arrigoni@tugraz.at}
\affiliation{Institute of Theoretical and Computational Physics, Graz University of Technology, 8010 Graz, Austria}

\date{\today}

\begin{abstract}
We study nonequilibrium thermoelectric transport properties of a
correlated impurity connected to two leads for temperatures below the
Kondo scale.  At finite bias, for which a current flows across the
leads, we investigate the differential response of the current to a
temperature gradient.  In particular, we compare the influence of a
bias voltage and of a finite temperature on this thermoelectric
response.  This is of interest from a fundamental point of view to
better understand the two different decoherence mechanisms produced by
a bias voltage and by temperature.  Our results show that in this
respect the thermoelectric response behaves differently from the
electric conductance.  In particular, while the latter displays a
similar qualitative behavior as a function of voltage and temperature,
both in theoretical and experimental investigations, qualitative
differences occur in the case of the thermoelectric response.  In
order to understand this effect, we analyze the different
contributions in connection to the behavior of the impurity spectral
function versus temperature.  Especially in the regime of strong
interactions and large enough bias voltages we obtain a simple picture
based on the asymmetric suppression or enhancement of the split Kondo
peaks as a function of the temperature gradient.  Besides the academic
interest, these studies could additionally provide valuable
information to assess the applicability of quantum dot devices as
responsive nanoscale temperature sensors.

\end{abstract}
\pacs{71.27.+a,72.15.Qm,73.63.Kv,73.23.-b}

\maketitle

\section{Introduction}
\label{sec:intro}

The Kondo effect~\cite{gl.re.88,ng.le.88,go.sh.98,cr.oo.98,sc.we.98} is one of the most prominent quantum many-body effects in nanoscopic physics. The (nonequilibrium) Kondo resonance~\cite{ro.pa.03,ke.05,sc.re.09}, being characterized by a much sharper characteristic line width $T_K$ than the resonant tunneling one determined by $\Gamma$ (s. ~\tcites{go.sh.98,cr.oo.98,si.bl.99,ma.so.96,de.sa.01,pa.ro.06,be.st.92}), allows to resolve electronic level splittings (e.g., in semiconductor nanostructures~\cite{zu.ma.04}, carbon nanotubes~\cite{sa.ja.05,hu.wi.09}, dopant atoms~\cite{zw.dz.13}), vibrational frequencies~\cite{os.on.07}, spin splittings due to a magnetic field~\cite{he.gu.04}, exchange interactions~\cite{de.sa.01,pa.ro.06,gr.ta.08}, magnetic anisotropies (e.g., in molecules~\cite{pa.ch.07,zy.va.10} or adatoms~\cite{lo.be.10}), and spin-orbit coupling~\cite{je.gr.11}.
While these have long been probed by gate controlled electrical transport spectroscopy, recent advancements in the experimental investigation of thermoelectric properties on the nanoscale~\cite{sc.bu.05,re.ja.07,ba.ma.08,wi.da.08,te.la.08,le.ki.13,ki.je.14} provide a promising route for the detection of additional information on the relaxation processes not accessible in the electrical transport~\cite{ge.ho.15}.
Theoretically, the development of powerful techniques led to significant progress in the understanding of correlated quantum dots out of equilibrium~\cite{co.gu.14,pr.gr.15,ha.co.15,gu.an.13,an.me.10,he.93,me.an.06,wi.me.94,le.sc.01,ro.kr.01,ko.sc.96,fu.ue.03,sm.gr.11,sm.gr.13,ju.li.12,sh.ro.05,fr.ke.10,ja.pl.10,sa.we.12,re.pl.14,an.sc.14,he.fe.09,ho.mc.09,nu.ga.13,nu.ga.15,an.08,sc.an.11,jo.an.13,we.ok.10,ha.he.07,nu.he.12,mu.bo.13,ro.pa.03,ke.05,pl.sc.12,do.nu.14,do.ga.15}.
An accurate description of the spectral and transport properties in the most challenging (nonperturbative) regime
of intermediate temperatures and bias voltages $T, \,\phi \lesssim T_K$ has however not been feasible until recently.
While the electronic transport has been studied extensively, the \emph{thermoelectric} transport theory mostly focused on the linear response regime~\cite{be.st.92,tu.ma.02,ko.vo.04,ku.ko.06,co.he.93,ki.he.02,sc.bu.05,co.zl.10,ku.ko.08,mu.me.08,an.co.11,re.zi.12,az.da.12,co.us.12,ro.to.12,ho.gh.13,ye.ho.14,sa.so.13,sv.ho.13,zi.ra.13,ke.sc.13}. The nonlinear regime has been addressed mainly in the weak coupling 
limit\cite{es.li.09,le.we.10,wa.wu.12,ke.sc.13,ga.ra.07,lo.sa.13,wh.13,fa.ho.13,zo.bu.14,ge.ho.15}, a systematic analysis including renormalization effects~\cite{pl.sc.12,kr.sh.12,ki.za.12,mu.bo.13} beyond the perturbative regime is still missing. 
Recent findings~\cite{le.we.10,ki.za.12} indicate a Seebeck coefficient which is enhanced with respect to the equilibrium result. A possible key to this observation are the different relaxation processes occurring at finite temperature and bias voltage. These differences in the behavior as a function of temperature and bias voltage offer a promising route to highly efficient devices~\cite{ha.ta.02,ma.04} operating in the nonlinear regime.
Besides quantum dot setups being considered as potential solid state energy converters~\cite{ma.so.96,be.st.92,ma.sa.97,gi.he.06,hu.ne.02,ki.he.02,sc.bu.05,ke.sc.13},
a variety of experimental realizations have been proposed, ranging from molecular systems~\cite{gi.st.87,jo.gi.00} to ultracold atoms, for which recent progress allows to address quantum transport in two terminal transport setups~\cite{th.we.99,br.me.12,br.gr.13,ch.pe.15,kr.st.15,kr.le.16}.
The high flexibility and control of these systems with highly tunable parameters provides a further interesting route to improve the understanding of the thermoelectric transport properties of strongly interacting systems.

In this work, we focus on the electronic contribution to the thermoelectric response of a quantum dot described by the single level Anderson model (SIAM) out of equilibrium. Aiming at understanding the microscopic processes underlying the physical behavior when applying a temperature difference in addition to a bias voltage, we investigate the differential response \djdT of the current to a temperature gradient rather than macroscopic thermoelectric properties such as the figure of merit or efficiency. In particular, we consider a finite bias voltage and an infinitesimal temperature difference \dDT and compute the differential response of the current \djdTns, which can be directly compared to equilibrium results. An extensive numerical renormalization group (NRG)~\cite{wi.75,bu.co.08} study of the thermoelectric properties of the SIAM in equilibrium can be found e.g. in \tcite{co.zl.10}. The differential quantities allow us to assess the effect of a finite temperature $T$ in presence of a finite bias voltage $\phi$. On the level of the spectral function, the equilibrium behavior for $\phi=0$ (and $T\neq0$) differs significantly from the nonequilibrium one for $\phi\neq0$. In equilibrium a single Kondo peak is suppressed upon increasing $T$, while in nonequilibrium the Kondo resonance splits into two weak excitations at the chemical potentials of each lead for values of $\phi$ of the order of $T_K$. Despite of these differences in the spectral function, the differential conductance \djdV exhibits  in the spin-symmetric case a qualitatively similar dependence  on $T$ and on $\phi$.
Here, we present calculations for \djdT as a function of $T$ or $\phi$ and for various system parameters, which is essentially the complementary information to \djdVns. We further analyze these results in connection with
the response of the nonequilibrium spectral function to a temperature gradient at finite bias.
Besides the fundamental interest, the knowledge of the behavior of \djdT as a function of the system parameters might be valuable for sensing applications since it determines the electric response to an applied temperature difference. 

We use the recent implementation of the auxiliary master equation approach (AMEA)~\cite{ar.kn.13,do.nu.14,ti.do.15,do.ti.16} within matrix product states (MPS)~\cite{do.ga.15}, which 
allows for a  significantly improved accuracy as compared to previous implementations based on exact diagonalization~\cite{ar.kn.13,do.nu.14}. The equilibrium results for the strongly interacting SIAM were benchmarked against NRG data, showing a remarkable agreement. In the AMEA results below, we obtained a high spectral resolution over the whole frequency domain for temperatures and voltages below {$k_{B}T_{K}/e$}.

Our paper is organized as follows: in the next section, we introduce the 
model and present the main concepts of the nonequilibrium impurity solver, the AMEA approach and its formulation in terms of Keldysh Green's functions. Details of the computation of differential quantities are specified in \app\ref{sec:fit}. In \se\ref{sec:results} we discuss our results for the thermoelectric transport properties at finite bias voltages. In particular, we analyze the differential response of the current to a temperature gradient in terms of the different contributions arising from the behavior of the spectral function as a function of temperature. We present a consistent physical picture in the nonequilibrium Kondo regime and conclude with a summary in \se\ref{sec:conclusio}.

\section{Model and Method}
\label{sec:modelmethod}
\subsection{Model}
\label{sec:model}
The nonequilibrium SIAM considered in this work is  given by a single site Hubbard model, 
which is connected two two leads $\lambda\in\{L/R\}$ at different chemical potentials $\mu_{L/R}$ and temperatures $T_{L/R}$. The corresponding Hamiltonian reads
\begin{equation}
  H =  H_{\mathrm{imp}} +  H_{\mathrm{leads}} +  H_{\mathrm{coup}}\,.
 \label{eq:H_parts}
\end{equation}
Here, $H_{\mathrm{imp}}$ describes the isolated impurity by
\begin{equation}
 H_{\mathrm{imp}} = \sum_{\sigma\in\{\uparrow,\downarrow\}} \varepsilon_f f^\dagger_{\sigma} f_{\sigma}^\nag + U n_{f\up}n_{f\dw}\,,
 \label{eq:H_imp}
\end{equation}
with the fermionic creation and annihilation operators for spin $\sigma$ denoted by $f^\dagger_\sigma$/$f^\nag_\sigma$ and $n_{f\sigma} = f^\dagger_{\sigma}f^\nag_{\sigma}$. The onsite Coulomb interaction is given by $U$ and the onsite energy for each spin by $\varepsilon_f = -U/2 + V_G$, such that the particle-hole symmetric point corresponds to a gate voltage of $V_G=0$. The whole Hamiltonian \eq{eq:H_parts} is assumed to be symmetric with respect to the spin degree of freedom. The Hamiltonian for the leads is given by
\begin{equation}
 H_{\mathrm{leads}} = \sum_{\lambda \in \{L,R\}} \sum_{k \sigma} \varepsilon_{\lambda k} d^\dagger_{\lambda k \sigma}d^\nag_{\lambda k \sigma} \,,
 \label{eq:H_res}
\end{equation}
with $d^\dagger_{\lambda k \sigma}$/$d^\nag_{\lambda k \sigma}$ the fermionic operators for lead electrons and $\varepsilon_{\lambda k}$ the energies of the $N$ lead levels  for each lead. The coupling Hamiltonian reads 
\begin{equation}
 H_{\mathrm{coup}} = \frac{1}{\sqrt{N}} \sum_{\lambda \in \{L,R\}} t'_\lambda  \sum_{ k \sigma} \left( d^\dagger_{\lambda k \sigma}f^\nag_{\sigma} + \mathrm{h.c.} \right) \,,
 \label{eq:H_coup}
\end{equation}
with coupling amplitudes $t'_\lambda$ which for simplicity we assume to be symmetric ($t'_L=t'_R$).
In particular, we choose for the leads a flatband model so that the lead density of states $\rho_\lambda(\omega)$ is constant in the range $(-D,D)$ and zero outside~\footnote{For AMEA, however, it is favorable to avoid discontinuities so that we introduce a smoothing of the edges at $\omega=\pm D$, cf. \tcite{do.ga.15}, which is irrelevant for large enough $D$.}.
We take the hybridization strength $\Gamma = \sum_\lambda\pi {t'_\lambda}^2\rho_\lambda(\omega=0)$ as our unit of energy and consider $D=10\,\Gamma$. The chemical potentials of the two leads are shifted anti-symmetrically by an external bias voltage $\phi$, i.e. $\mu_L = - \phi/2$ and $\mu_R = + \phi/2$, and also a temperature difference \DT is applied in the same manner $T_L = T - \dT/2$ and $T_R = T + \dT/2$, with $T$ the average temperature.

In the equilibrium limit $\phi=0$ and $\dT=0$ the low-energy physics of the SIAM is governed by the Kondo scale $T_K$. It becomes exponentially small for large values of the interaction since $T_K \propto \exp(-\pi U / 8 \Gamma)$ for $V_G=0$\cite{he.97}.
We here define $T_K$ by $G(T=T_K) = G_0/2$ (at $V_G=0$), with $G(T)$ the temperature-dependent conductivity and $G_0$ the quantum limit obtained for $T\to0$~\cite{bu.co.08,os.zi.13}. 
This choice of $T_K$ directly related to observables is especially suited for comparison with experiments. From a NRG calculation~\footnoteremember{noteNRG}{
All numerical renormalization group (NRG) calculations were performed with the open source code "NRG Ljubljana", see \url{http://nrgljubljana.ijs.si/}.}
we find for $V_G=0$ and interaction strengths of $U=4\,\Gamma$, $U=6\,\Gamma$ and $U=8\,\Gamma$ values of $T_K = 0.50\,\Gamma$, $T_K = 0.21\,\Gamma$ and $T_K = 0.10\,\Gamma$, 
respectively.
Away from particle-hole symmetry and for fixed $U$, $T_K$ increases as a function of $V_G$ since
 $T_K \propto \exp(-\pi (U^2-4V_G^2) / (8 \Gamma U))$~\cite{he.97,co.16}.
For the calculations presented below we consider $V_G>0$ and a fixed temperature of $T=0.1\,\Gamma$, leading to the regime $T<T_K$.

\subsection{Keldysh Green's functions}
\label{sec:Keldysh}
The nonequilibrium SIAM introduced in the previous section is conveniently addressed in the framework of Keldysh Green's functions~\cite{sc.61,ka.ba.62,ke.65,ha.ja.98,ra.sm.86,st.le.13}. In the steady state limit one needs to consider an independent retarded $G^R(\omega)$ and Keldysh Green's function $G^K(\omega)$, which are defined in the time domain by
\begin{align}
 G^R(t) &= -i \Theta(t) \Braket{ \left\{ c(t),  c^\dagger \right\}  } \,, \nonumber \\
 G^K(t) &= -i  \Braket{ \left[ c(t),  c^\dagger \right]  } \,,
 \label{eq:GFs_t}
\end{align}
with $\{ A, B \}$ the anti- and $[ A, B ]$ the commutator of $A$ and $B$, and $\Theta(t)$ the Heaviside step function. 
The Green's functions in frequency domain are obtained as usual by Fourier transformation $G(\omega) = \int G(t) \exp(i\omega t)dt$. In equilibrium $G^R(\omega)$ and $G^K(\omega)$ are related to each other via the fluctuation-dissipation theorem
\begin{equation}
G^K(\omega) = 2\pi i\left(2 f(\omega,\mu,T) -1 \right) A(\omega)\,,
 \label{eq:FDT}
\end{equation}
with $f(\omega,\mu,T)$ the Fermi function for temperature $T$ and chemical potential $\mu$, the spectral function
\begin{align}
 A(\omega) = \frac{i}{2\pi} \left( G^R(\omega) - G^A(\omega) \right) \,,
 \label{eq:Aw}
\end{align}
and the advanced Green's function $G^A(\omega) = G^R(\omega)^\dagger$.

It is common to adopt a $2\times2$ matrix structure for Green's functions
\begin{equation}
   \und{G}(\omega) = \begin{pmatrix} G^R(\omega) & G^K(\omega) \\ 0 & G^A(\omega) \end{pmatrix}\,,
  \label{eq:negf}
\end{equation}
and other
 two-point functions. We will denote such 
$2\times2$ objects by
an underscore $\und{\dots}$.
Products of such objects are conveniently evaluated with the Langreth rules, e.g. for the inverse one finds~\cite{ha.ja.98}
\begin{align}
 \left(\und{G}^{-1}\right)^R &=  \left({G}^R\right)^{-1} \,,\nonumber \\
 \left(\und{G}^{-1}\right)^K &=  - \left({G}^R\right)^{-1} {G}^K \left({G}^A\right)^{-1} \,.
 \label{eq:invKeldysh}
\end{align}
With this, Dyson's equation for the SIAM can be written as
\begin{align}
\und{G}^{-1}(\omega) &=\und{g}^{-1}_{0}(\omega) - \und{\Delta}(\omega)-\und{\Sigma}(\omega)\nonumber \\
		  &=\und{G}^{-1}_{0}(\omega) -\und{\Sigma}(\omega)\,,
\label{eq:dyson}
\end{align}
where $\und{g}^{-1}_{0}(\omega)$ is the noninteracting Green's function of the isolated impurity \footnote{As usual, $g^R_{0}(\omega) = (\omega - \varepsilon_f +i0^+)^{-1}$ and the contribution from $g^K_{0}(\omega)$ is infinitesimal in the steady state.}, $\und{\Delta}(\omega)$ the hybridization function accounting for $H_{\mathrm{leads}} +  H_{\mathrm{coup}}$, $\und{\Sigma}(\omega)$ the self-energy for the interacting coupled system, and $\und{G}^{-1}_{0}(\omega)$ the noninteracting Green's function of the impurity coupled to leads. As usual for many-body problems, the calculation of  
$\und{\Sigma}(\omega)$ is most demanding and cannot be done exactly in general. 
The main concepts of the nonequilibrium impurity solver used in the present work are specified in \se\ref{sec:method}, for more details we refer to~\tcites{ar.kn.13,do.nu.14,do.ga.15}. The hybridization can be expressed as
\begin{equation}
 \und{\Delta}(\omega) = \sum_\lambda {t'_\lambda}^2 \und{g}_\lambda(\omega) \,,
 \label{eq:Dhyb}
\end{equation} 
in terms of
the {boundary} leads' Green's functions $\und{g}_\lambda(\omega)$, whose retarded component is related to the density of states by $\rho_\lambda(\omega) = -1/\pi\iim\{g^R_\lambda(\omega) \}$. Its Keldysh component is determined by observing that in the decoupled case, i.e. at infinite past on the Keldysh contour, the leads are in equilibrium so that the retarded and Keldysh part of $\und{g}_\lambda(\omega)$  fulfill \eq{eq:FDT}.

\subsection{Thermoelectric response}
\label{sec:thermo}
As stated above, in this work we are interested in computing the thermoelectric properties for a finite bias voltage $\phi\neq0$ and an infinitesimal temperature difference \dDTns.
In the framework of Keldysh Green's functions the current across the impurity can be calculated with the aid of the Meir-Wingreen expression~\cite{me.wi.92}. In our case, since we consider a bias-independent lead density of states $\rho_L(\omega)=\rho_R(\omega)$, a simplified Landauer-type formula is obtained~\cite{ha.ja.98}
\begin{equation}
 j = \int \mathcal T(\omega) \big(f_R(\omega) - f_L(\omega) \big) d\omega\,,
 \label{eq:current}
\end{equation}
with the Fermi functions for the leads
 $f_\lambda(\omega) = 1 / (\exp[(\omega-\mu_\lambda)/T_\lambda] + 1 )$
and the transmission given by $\mathcal T(\omega) = A(\omega)\gamma(\omega)$, where $\gamma(\omega) = \sum_\lambda\pi {t'_\lambda}^2\rho_\lambda(\omega)$.
For the differential quantities below we additionally specify the derivative $f_\lambda'(\omega) = -1/(4\cosh^2[(\omega-\mu_\lambda)/2T_\lambda])$, which is a peaked function around $\omega=\mu_\lambda$.
Similarly to the electric current, the heat current at the left- or right-sided lead can be calculated by replacing $\mathcal T(\omega) \to (\omega-\mu_\lambda)\mathcal T(\omega)$ in \eq{eq:current} above~\cite{co.zl.10,ki.za.12}. Notice that, in contrast to the particle and energy current, the nonequilibrium heat current is not conserved and the left and right contributions differ~\cite{ki.za.12}.

Here, however, we focus on the electric current and its response to an infinitesimal change in the temperature difference or to a voltage
\newcommand{\cals}{{\cal S}}
\emph{in nonequilibrium,} i.e. at finite $\phi$. This prompts us to introduce a nonequilibrium generalization of the linear response Seebeck coefficient $\cals$ to the case of a nonzero current $j$ as
\begin{equation}
  \cals := - \left.\dfrac{d\phi}{d(\dT)}\right|_{j=\mathrm{const.}} =  \left.{\dfrac{\partial j}{\partial(\dT)}}  \,\,\middle/\,\,   {\dfrac{\partial j}{\partial \phi}} \right. \,.
 \label{eq:dS}
\end{equation}
We expect $\cals$ to be well accessible by experiments since it may be measured through a temporal modulation of \DT at constant current. With $\cals$ and the knowledge of \djdV the thermoelectric response \djdTns, which is the object of our further analysis, can be determined.

To evaluate \djdT we differentiate \eq{eq:current} with respect to \DT and find two contributions
\begin{equation}
 {\dfrac{\partial j}{\partial(\dT)}} = {\dfrac{\partial j}{\partial(\dT)}}_A + {\dfrac{\partial j}{\partial(\dT)}}_B\,,
 \label{eq:djdT}
\end{equation}
with
\begin{equation}
 {\dfrac{\partial j}{\partial(\dT)}}_A  \hspace{-1ex}= \int  \dfrac{\partial \mathcal T(\omega)}{\partial(\dT)}  \big(f_R(\omega) - f_L(\omega) \big) d\omega\,,
  \label{eq:djdT_A}
\end{equation}
and 
\begin{equation}
 {\dfrac{\partial j}{\partial(\dT)}}_B  \hspace{-1ex}= \frac{1}{2T^2} \int  \mathcal T(\omega) \sum_\lambda -f'_\lambda(\omega)(\omega-\mu_\lambda) d\omega\,.
  \label{eq:djdT_B}
\end{equation}
In equilibrium, \eq{eq:djdT_A} vanishes and \eq{eq:djdT_B} reduces to the well-known linear response expression. In nonequilibrium, however, the situation is more involved and $\partial \mathcal T(\omega)/\partial(\dT)$ has to be determined numerically. This is addressed in more detail in \app\ref{sec:fit}. 
The second term \eq{eq:djdT_B}, in contrast, is rather intuitive since it is given by the asymmetry of $\mathcal T(\omega)$ at $\omega = \mu_{L/R}$. 
As in equilibrium, \eq{eq:djdT_B}, and also \eq{eq:djdT}, vanishes in the limit of particle-hole symmetry, i.e. $V_G=0$~\footnote{\eq{eq:djdT_A} vanishes for a particle-hole symmetric situation since $\partial \mathcal T / \partial (\Delta T)$ is anti-symmetric with respect to $\omega$. The reason is that the hybridization function for $\dT>0$, and thus also $\mathcal T(\omega)$, is related to the one for $\dT<0$ by a particle-hole transformation.}.
We finally note that in the present analysis we will focus on evaluating \eq{eq:djdT} for cases with $\phi\neq0$ and $\dT=0$.

\subsection{Method}
\label{sec:method}
\begin{figure*}
\begin{center}
\includegraphics[width=0.33\textwidth]{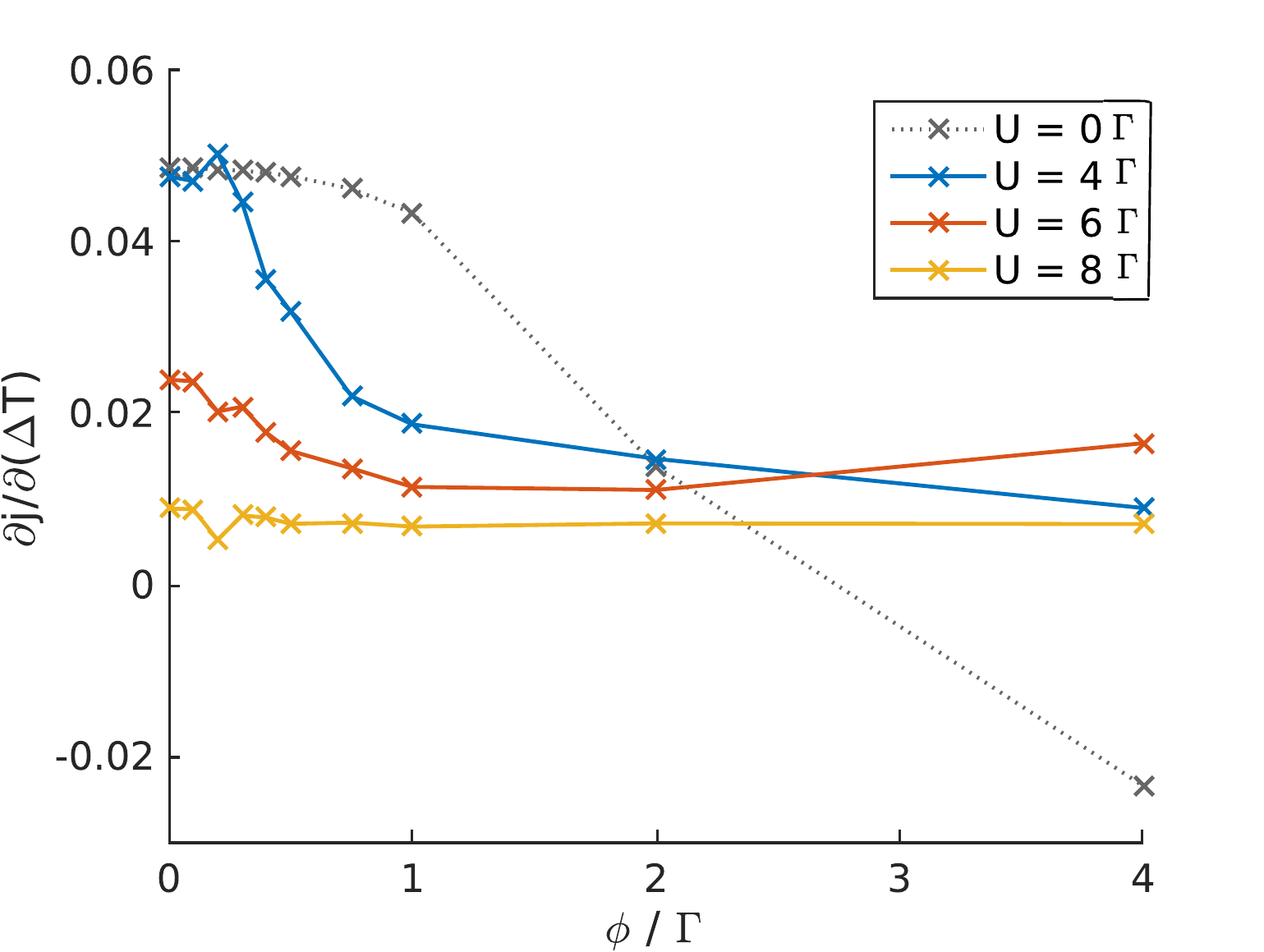}\hspace{-0.02\textwidth}
\includegraphics[width=0.33\textwidth]{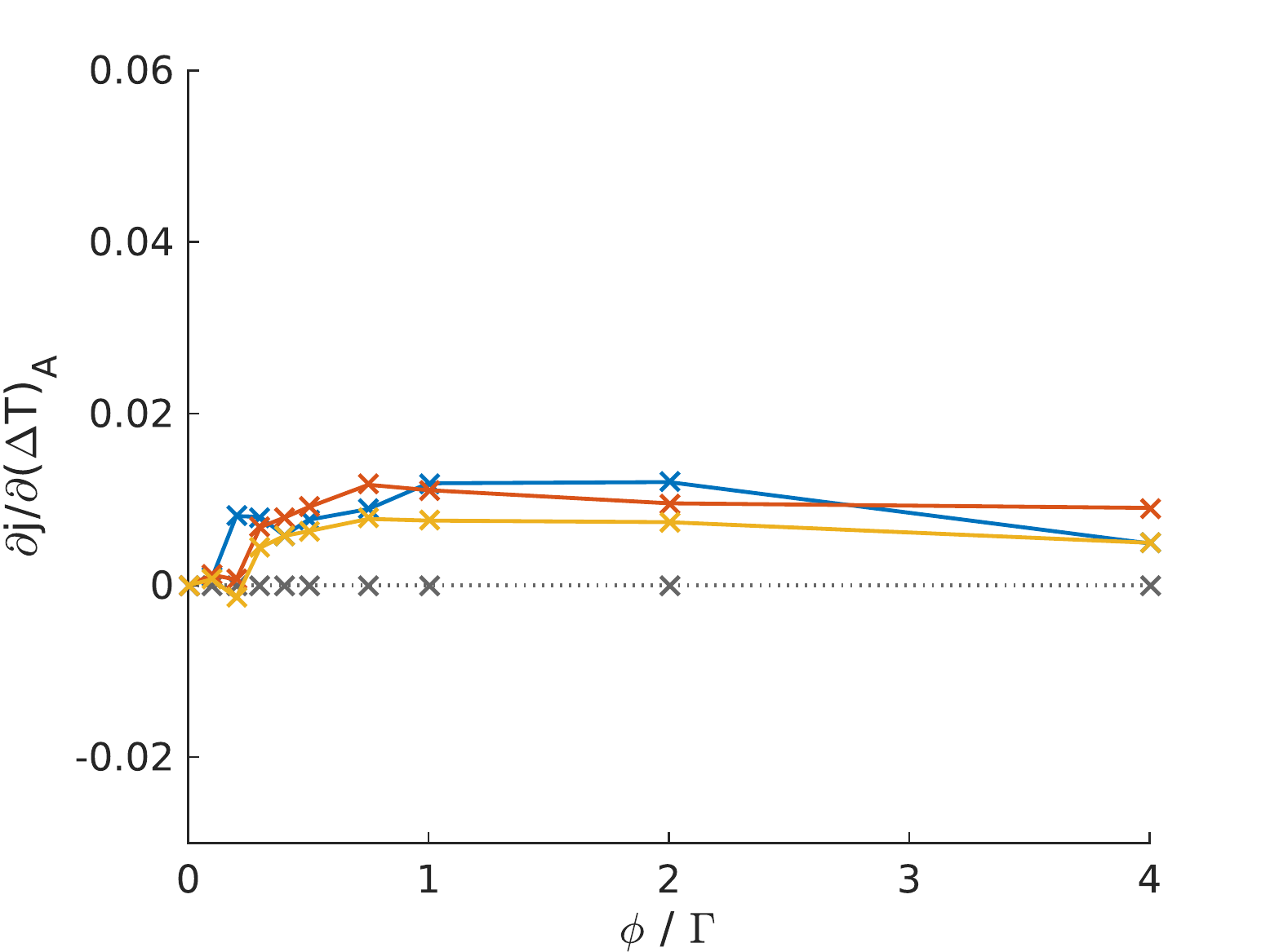}\hspace{-0.02\textwidth}
\includegraphics[width=0.33\textwidth]{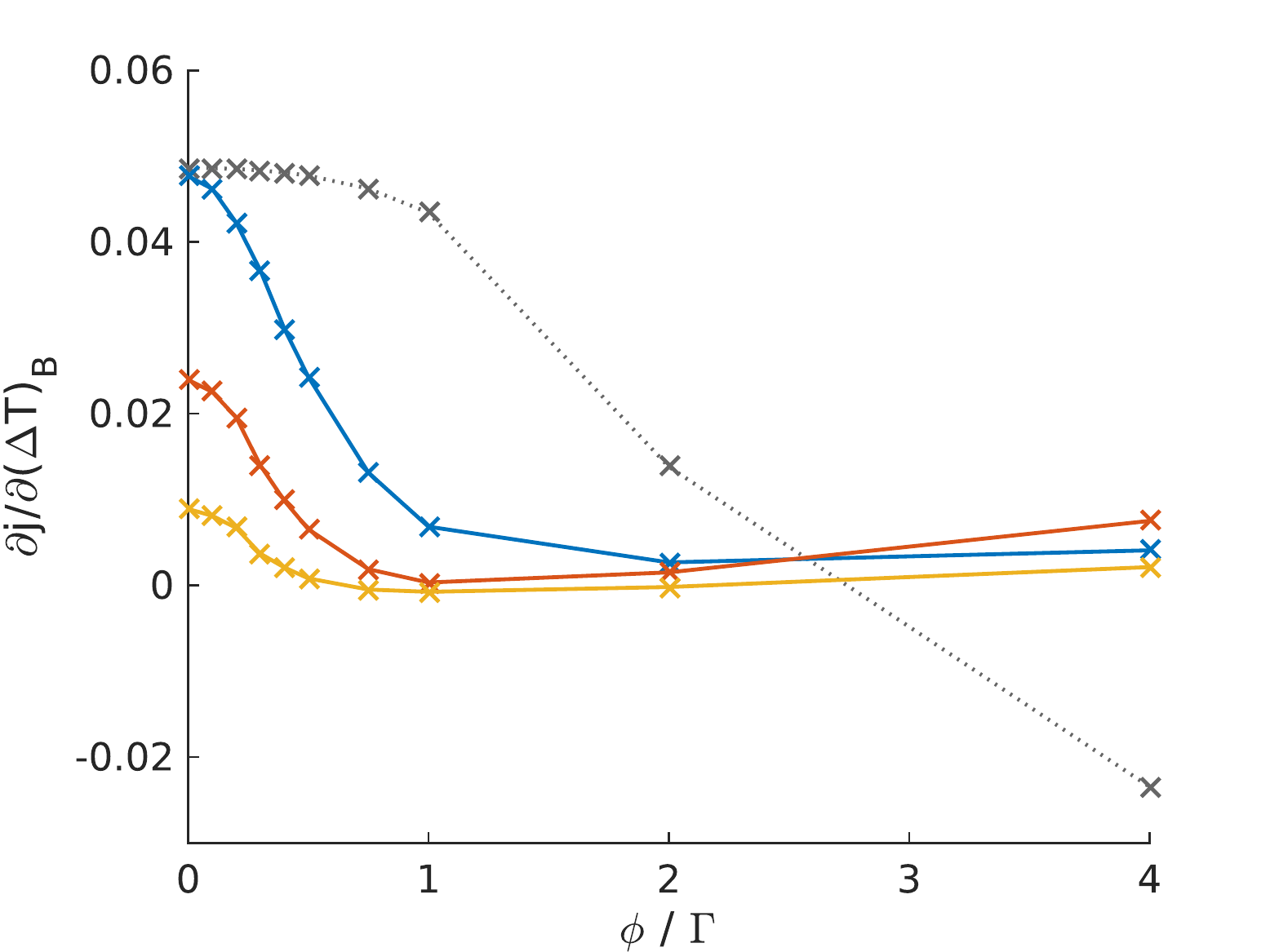}
\caption{(Color online) Thermoelectric response for $V_G = \Gamma$ and for various values of the interaction strength $U$. The first panel depicts \djdTns, \eq{eq:djdT}, and the second and third panel the two parts of \eqs{eq:djdT_A} and (\ref{eq:djdT_B}). See App.~\ref{sec:fit} for a discussion on the accuracy of the numerical derivative.
}
\label{fig:djdT_VG1}
\end{center}
\end{figure*}
In order to determine the nonequilibrium self-energy $\und{\Sigma}(\omega)$ we here use the so-called auxiliary master equation approach (AMEA), as previously introduced in \tcites{ar.kn.13,do.nu.14} and in its recent MPS implementation \tcite{do.ga.15}. The basic principle of the approach is to map the original nonequilibrium (or equilibrium) impurity problem onto an auxiliary one, in which the bath is modeled by a small number $N_B$ of bath sites and additional Markovian environments. By this, a finite open quantum system described by a Lindblad equation is obtained, which can be solved accurately by numerical techniques. The mapping procedure relies on the condition to represent the original hybridization function as accurately as possible by the auxiliary one. A similar idea is commonly used in the context of dynamical mean field theory~\cite{me.vo.89,ge.ko.92}. However, in order to be able to address nonequilibrium situations it is necessary to fit not only the retarded but also the Keldysh component, and therefore to construct an auxiliary hybridization with a continuous spectrum. The AMEA approach is conveniently formulated in terms of Keldysh Green's functions~\cite{sc.61,ka.ba.62,ke.65,ha.ja.98,ra.sm.86,st.le.13}, and particularly suited to treat steady state situations. Time-dependent problems are also accessible in principle, but go beyond the scope of the present work.

AMEA provides an accurate approximation for the impurity self-energy $\und{\Sigma}(\omega)$, for details on the method we refer to \tcite{do.ga.15}. The accuracy is hereby controlled by the quality of the mapping procedure and can be systematically improved by increasing $N_B$, the convergence being exponential in $N_B$~\cite{do.ga.15,do.so.16u}. In practice, of course, finite values for $N_B$ have to be chosen and with that techniques we are currently able to consider as many as $N_B \approx 10-20$. The calculations in \tcite{do.ga.15} demonstrated that these values allow for very accurate results. All of the calculations presented below are for $N_B=14$.

For the calculation of thermoelectric properties one needs to consider systems away from particle-hole symmetry. 
In our previous calculations in \tcite{ar.kn.13,do.nu.14,do.ga.15} we treated particle-hole symmetric situations only. However, this was done for convenience and is by no means a restriction of the method. The difference is that for $V_G \neq 0$ two independent Green's functions have to be computed in the many-body solution and that twice as many fit parameters must be considered in the mapping procedure.
Both aspects require increased computational resources, mainly due to the many-body calculation, but no essential conceptual modifications. The differential quantities \djdT are estimated by finite differences, see \app\ref{sec:fit} for more details on the implementation.

\section{Results}
\label{sec:results}
\begin{figure*}
\begin{center}
\includegraphics[width=0.33\textwidth]{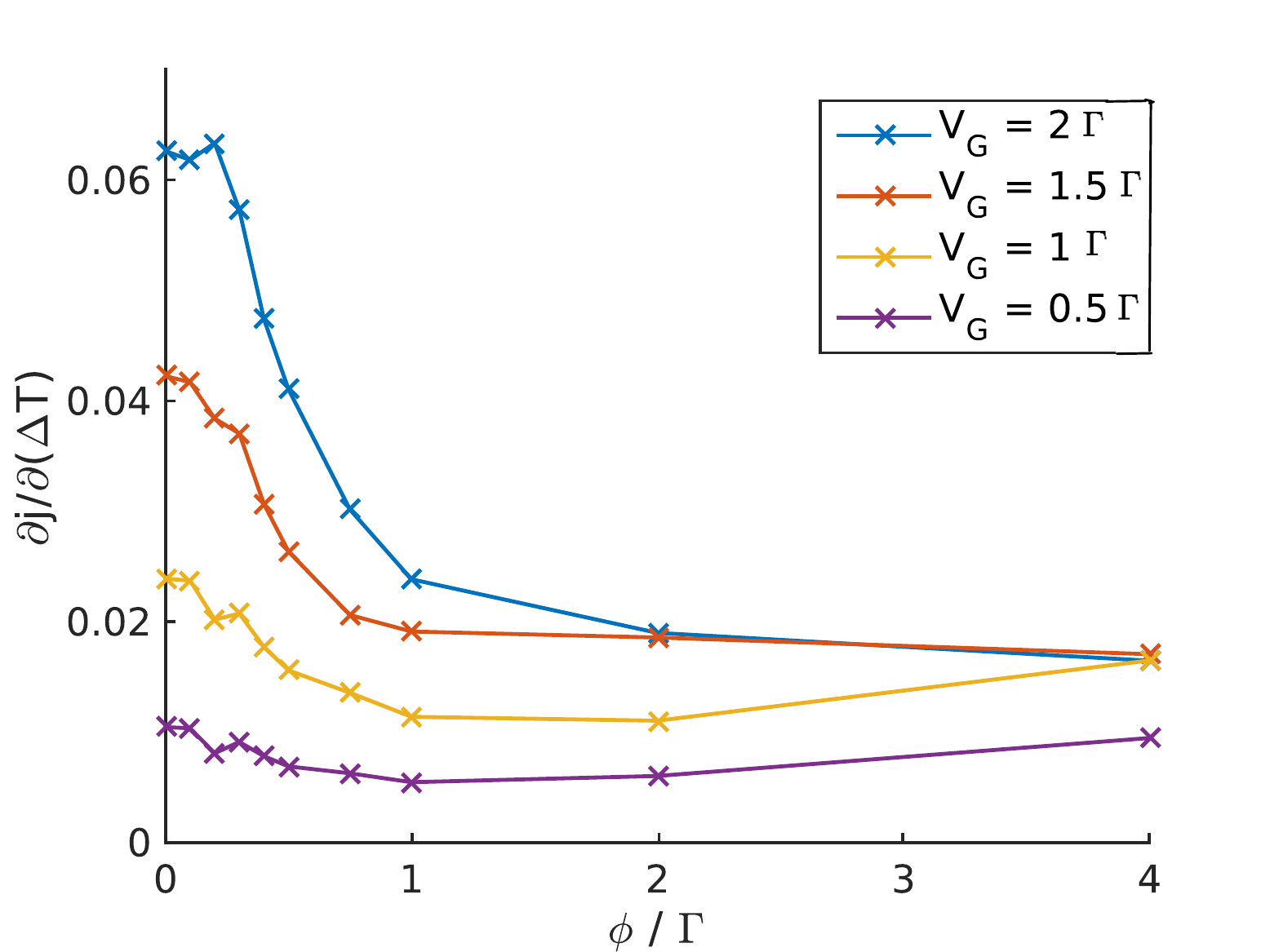}\hspace{-0.02\textwidth}
\includegraphics[width=0.33\textwidth]{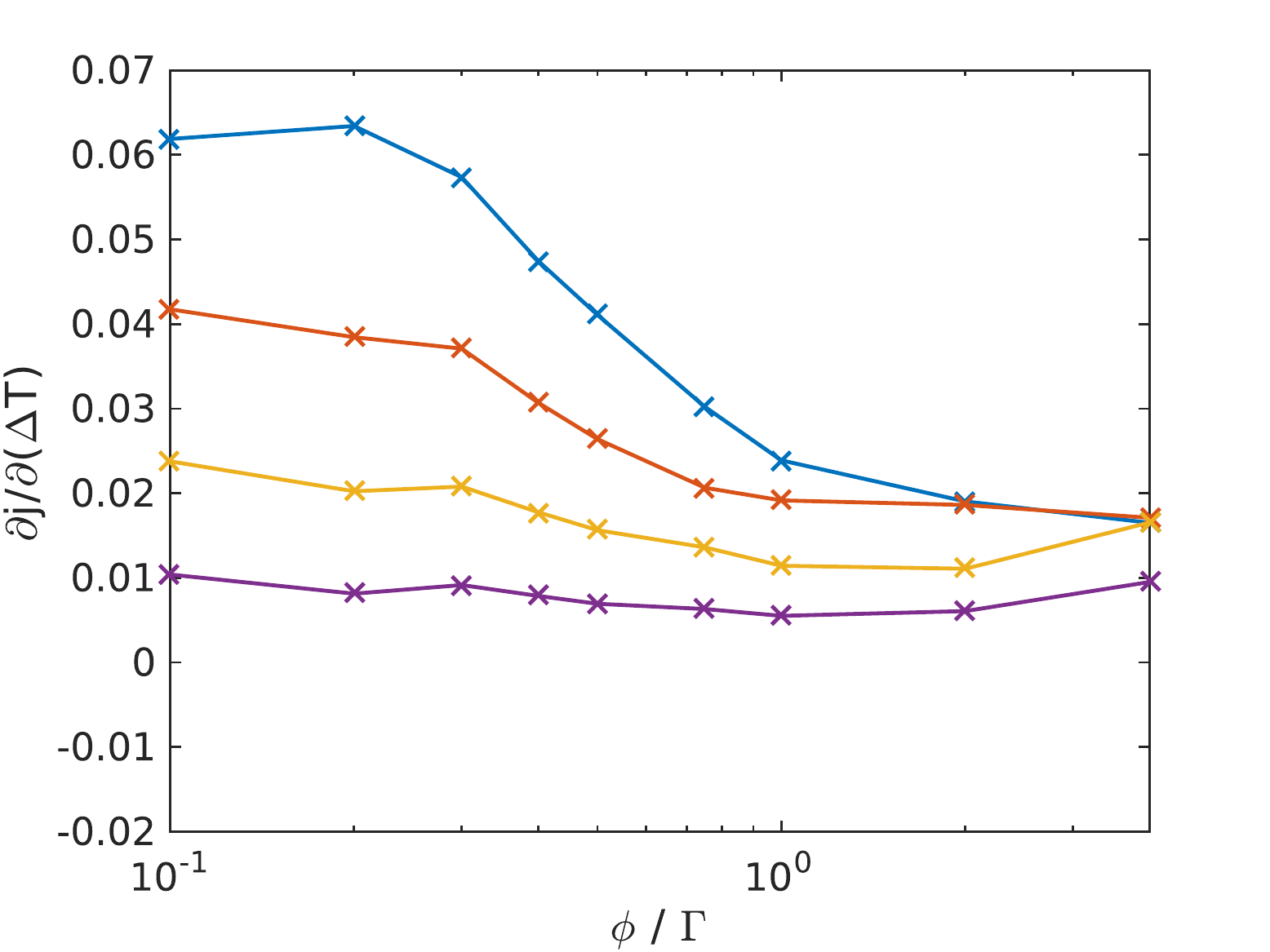}\hspace{-0.02\textwidth}
\includegraphics[width=0.33\textwidth]{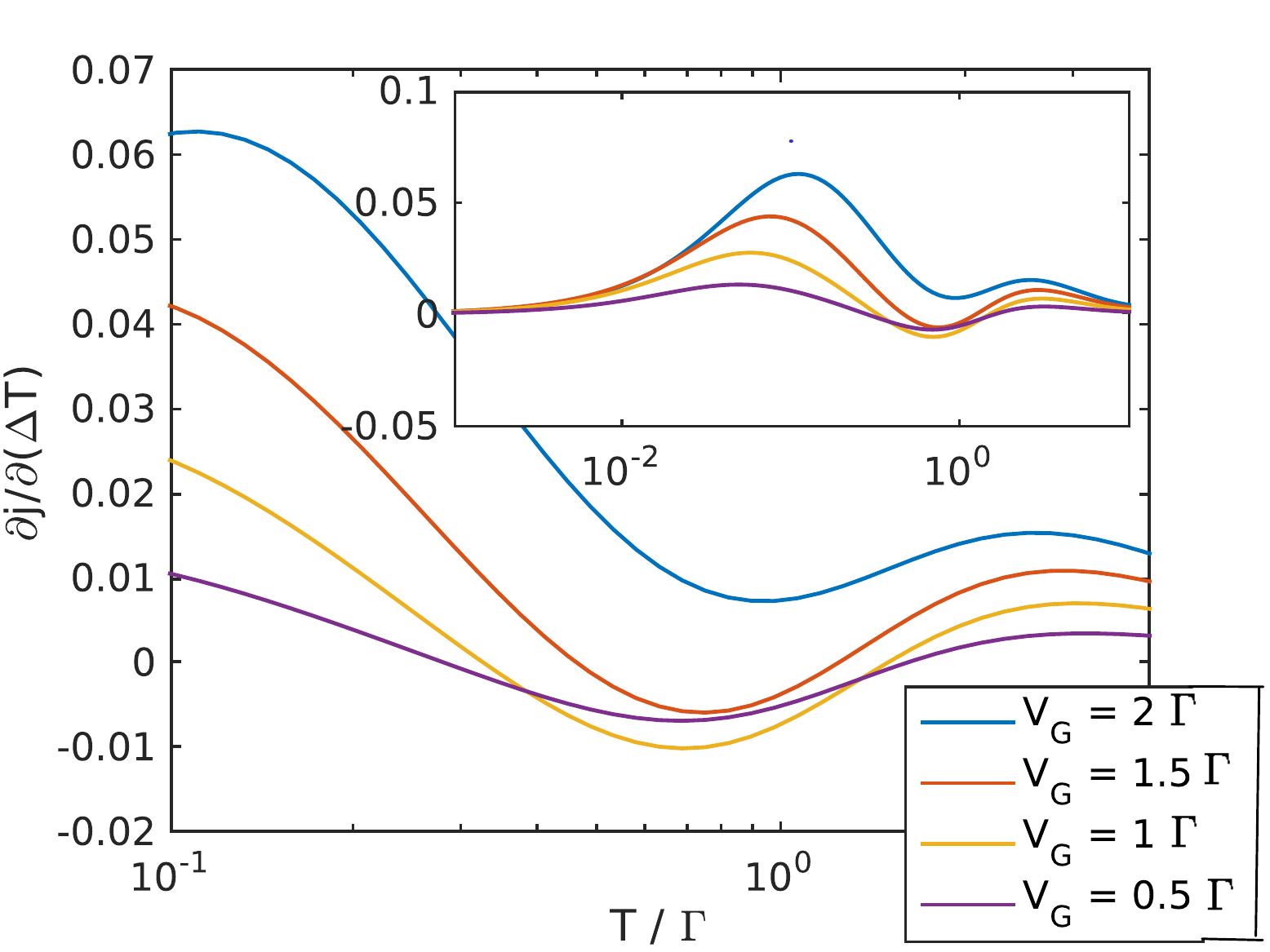}
\caption{(Color online) Thermoelectric response \djdT as a function of bias voltage $\phi$ in the first panel (in the second on a semi-logarithmic scale), and in the third for $\phi=0$ as a function of $T$. All calculations are for $U=6\,\Gamma$ and for various $V_G$. 
The equilibrium values on the right are obtained from NRG~\footnoterecall{noteNRG}.
The inset depicts the behavior for a larger temperature range.
}
\label{fig:djdT_U6}
\end{center}
\end{figure*}

We start by investigating the current response \djdT of the nonequilibrium SIAM to an infinitesimal temperature difference \dDTns. We are interested, in particular, in the nonequilibrium behavior as a function of  bias voltage $\phi$. For all calculations we used an  average temperature $T=0.1\,\Gamma$, which, for $U=6\,\Gamma$ and $V_G\approx2\,\Gamma$, corresponds to the point at which \djdT as function of $T$ has a maximum below $T_K$. This is determined from an equilibrium NRG calculation~\footnoterecall{noteNRG}, see inset of the third panel of \fig{fig:djdT_U6}.

In \fig{fig:djdT_VG1} we analyze the general behavior of \djdT as a function of $\phi$ as determined by \eq{eq:djdT} together with its two contributions \eqs{eq:djdT_A} and (\ref{eq:djdT_B}), for different values of $U$ at a gate voltage $V_G=\Gamma$. In the noninteracting case, a plateau up to $\phi = V_G$ precedes the decrease of \djdT as a function of $\phi$~\footnote{This was checked for other $V_G$ as well, which showed that \djdT develops a peak at $\phi = V_G$ for $V_G>\Gamma$.}.
In this case only the second term \mbox{\djdTns$_B$} contributes since the lead density of states, and thus the noninteracting spectral function $A(\omega)$, is independent of $\phi$ and $T$. As a consequence, according to \eq{eq:djdT_B}, for $U=0$ we have that \djdT just measures the asymmetry of the equilibrium $A(\omega)$ at $\omega=\pm\phi/2$. 

For a finite value of $U\geq4\,\Gamma$ in the Kondo regime, the results change significantly. The current response \djdT is positive for all $\phi$ and decreases much faster as a function of $\phi$, with a dependence that resembles the one of \djdVns, see e.g. \tcites{do.ga.15,pl.sc.12,an.08,ha.he.07,co.gu.14,ro.pa.03,ke.05,sc.re.09}. However, in contrast to the conductance, in this case the relevant energy scale controlling the decrease can not be identified with 
$T_K$. In particular, it does not depend on $U$ and appears to be proportional to $\Gamma$. This agrees with the findings in \tcite{co.zl.10} and with the results below. We note the functional behavior of the two contributions \mbox{\djdTns$_A$} and \mbox{\djdTns$_B$} is very different, as shown in the second and third panel of \fig{fig:djdT_VG1}.
While \mbox{\djdTns$_B$} decreases with $\phi$ (being maximal for $\phi=0$) and strongly depends on the interaction, \mbox{\djdTns$_A$} in contrast, exhibits a similar behavior for all values of $U$: starting from zero, \mbox{\djdTns$_A$} increases with applied bias and saturates at $\phi\approx\Gamma$. For $U=8\,\Gamma$ the two contributions compensate each other yielding a nearly $\phi$-independent \djdTns.

\begin{figure*}
\begin{center}
\includegraphics[width=0.34\textwidth]{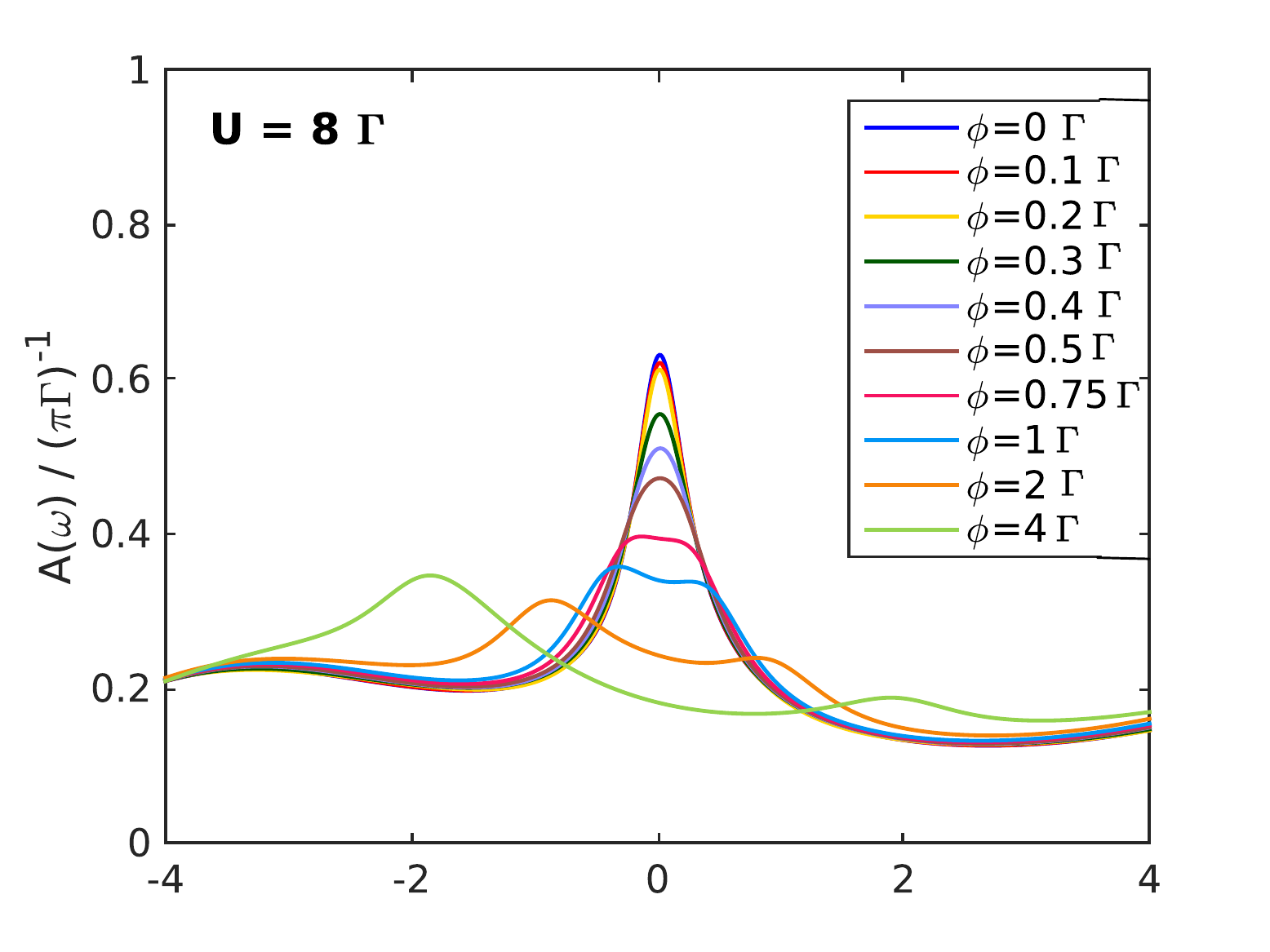}\hspace{-0.03\textwidth}
\includegraphics[width=0.34\textwidth]{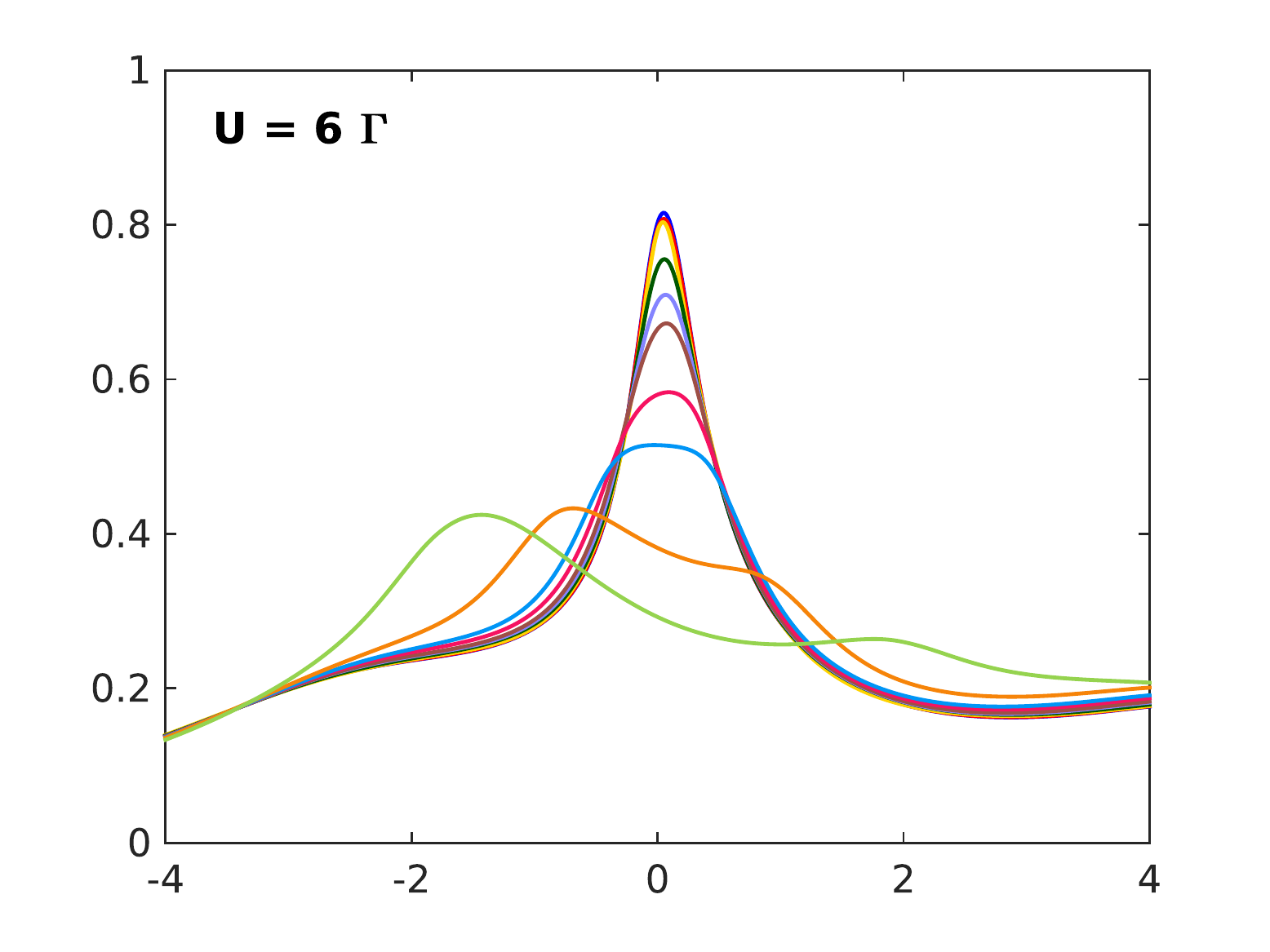}\hspace{-0.03\textwidth}
\includegraphics[width=0.34\textwidth]{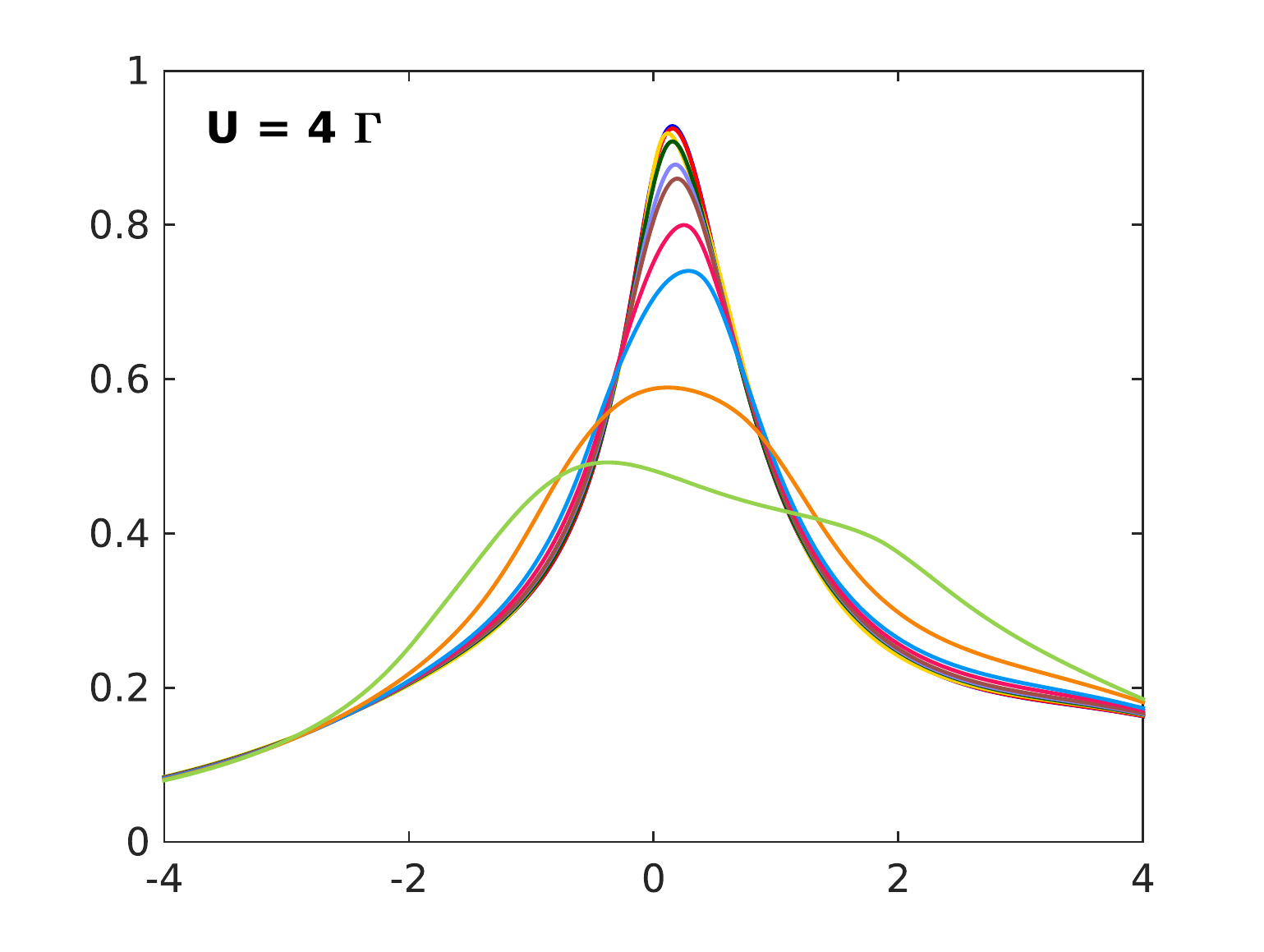}\\
\includegraphics[width=0.34\textwidth]{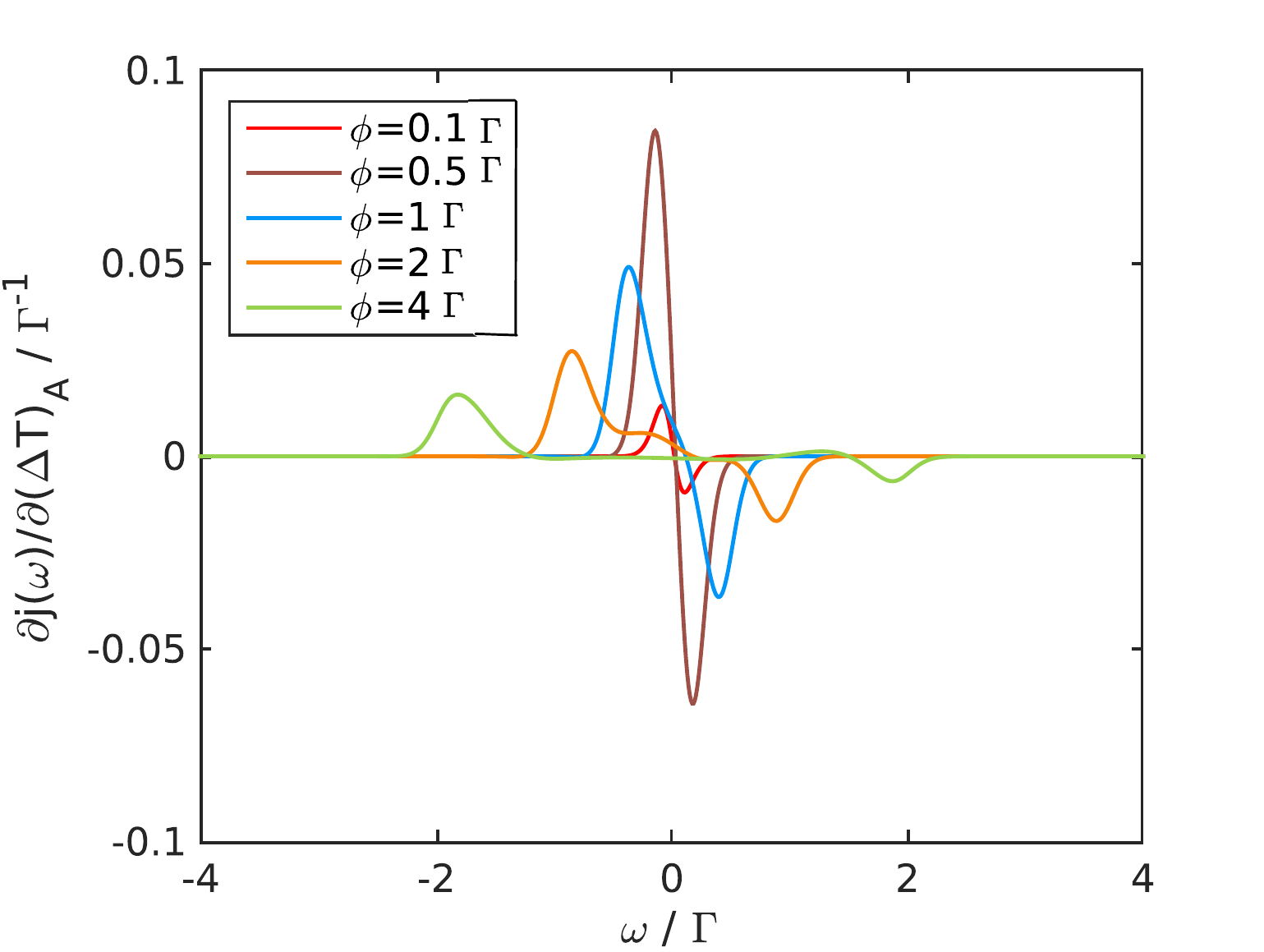}\hspace{-0.03\textwidth}
\includegraphics[width=0.34\textwidth]{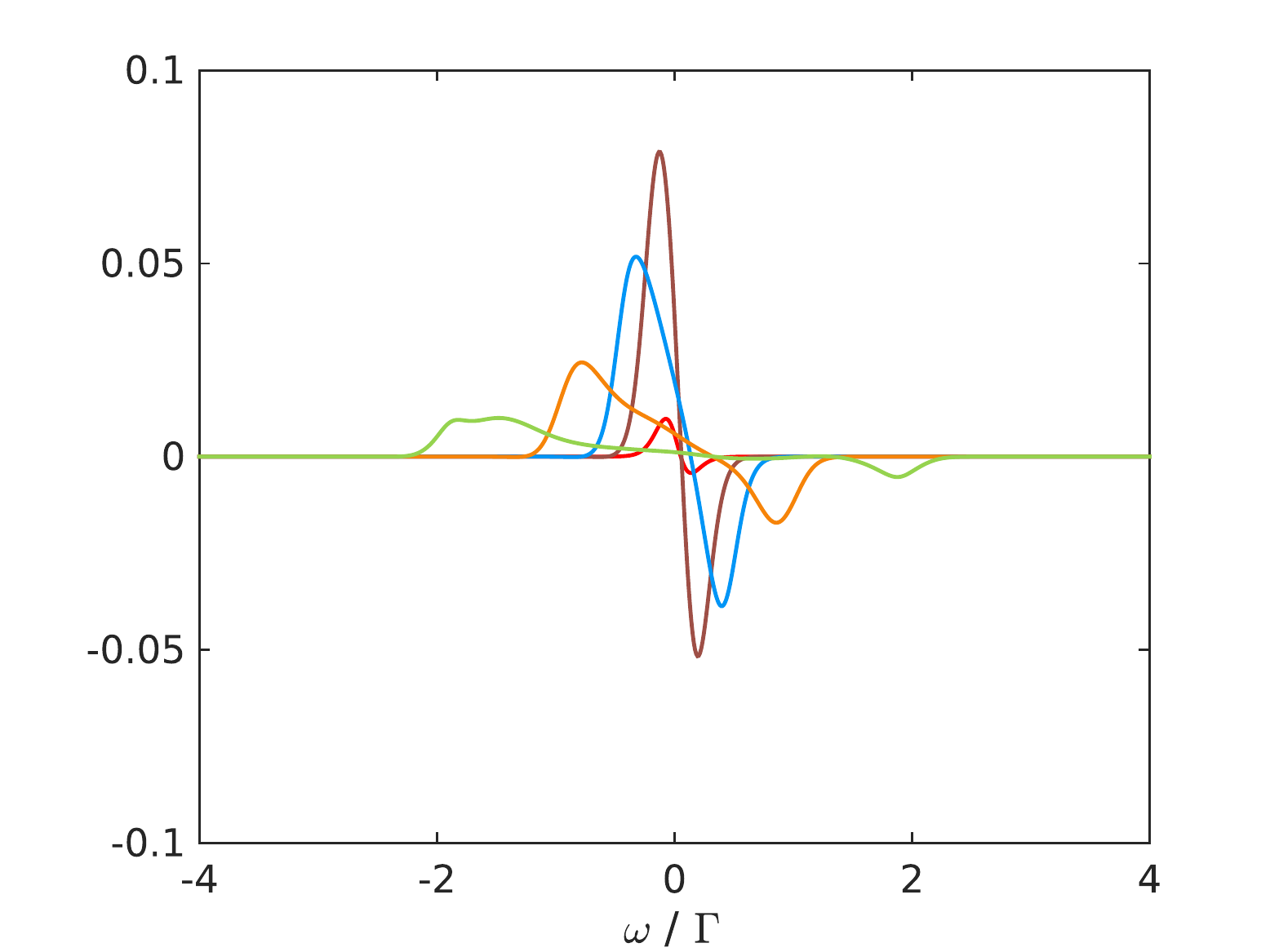}\hspace{-0.03\textwidth}
\includegraphics[width=0.34\textwidth]{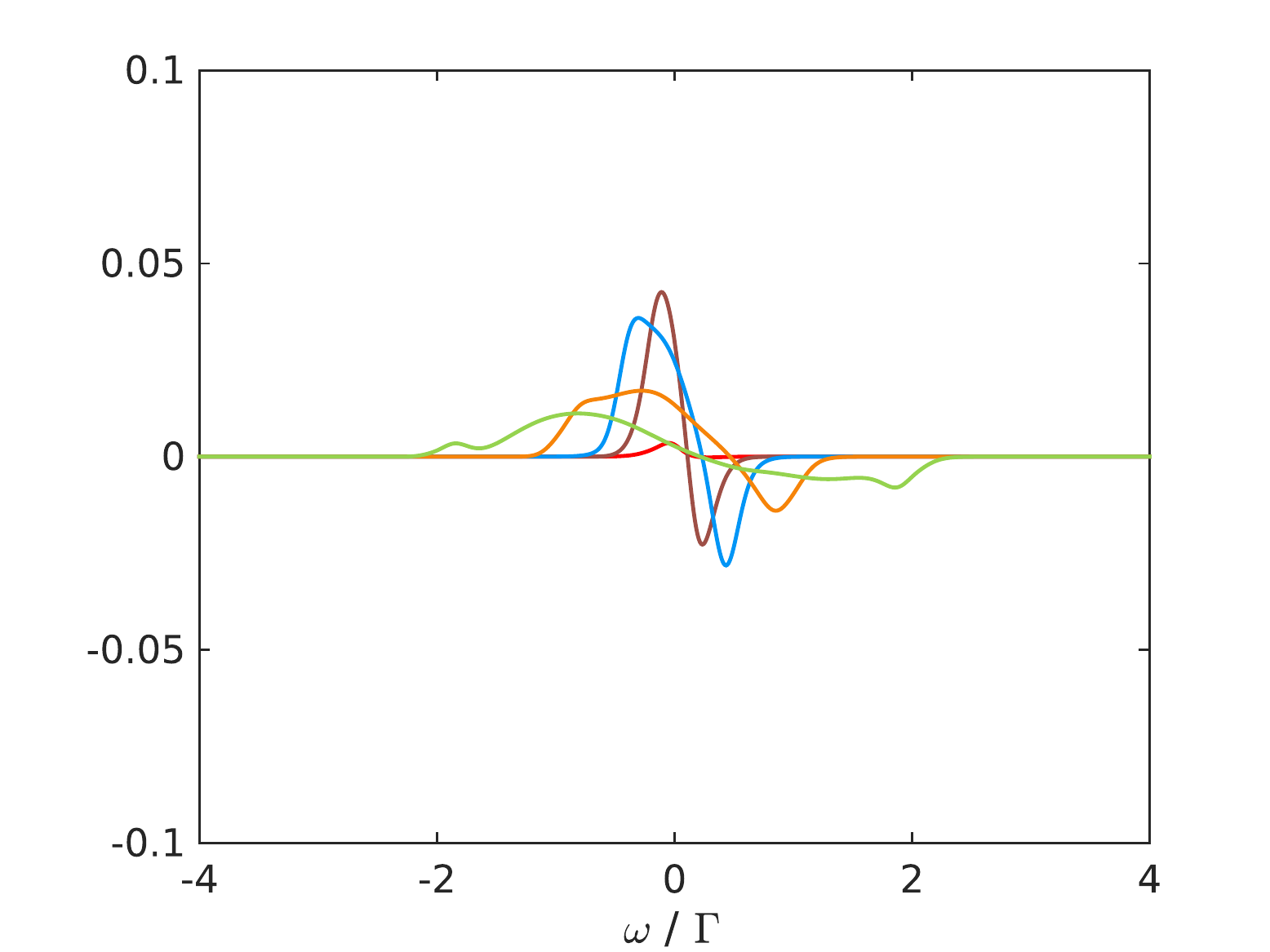}
\caption{(Color online) Spectral functions $A(\omega)$ for $\dT=0$ on top and of the integrand of \eq{eq:djdT_A} $\partial j(\omega) / \partial (\dT)$ at the bottom, which is proportional to $\partial A(\omega) / \partial (\dT)$. The results are shown for the same parameters as in \fig{fig:djdT_VG1}, i.e. on the left for $U=8\,\Gamma$, in the center for $U=6\,\Gamma$ and on the right for $U=4\,\Gamma$, with $V_G=\Gamma$ in all cases.
}
\label{fig:spectral_djdT_VG1}
\end{center}
\end{figure*}
In order to gain a more detailed understanding of this behavior we analyze the frequency dependence of the spectral function as well as of  the integrand determining \djdTns. The nonequilibrium spectral functions for $\dT=0$, which determine the second contribution \mbox{\djdTns$_B$}, are shown in the upper panels of \fig{fig:spectral_djdT_VG1}. We observe a pronounced Kondo peak at $\phi=0$ for all values of $U$. {For the results depicted in \fig{fig:spectral_djdT_VG1} the 
 temperature is fixed to $T=0.1\,\Gamma$ and $U$ decreases from left to right. Hence the ratio
$T/T_K$ decreases from left to right, resulting in an increasing height of the Kondo peak. }{Moreover,  it appears that for the smallest value of the interaction the peak is rather broad as it is not well separated from the lower Hubbard band}~\footnoteremember{mixedValence}{Note that the SIAM with $\varepsilon_f = V_G - U/2 = -\Gamma$ is already in the transition to the mixed valence regime.} as compared to larger $U$. 
The low-energy details are less smeared out for $U\geq 6\,\Gamma$ for which a splitting of the Kondo resonance into two peaks is clearly visible for sufficiently large $\phi$. In contrast to the particle-hole symmetric case (s. \tcite{do.ga.15}) the splitting is asymmetric in presence of a finite gate voltage.
According to \eq{eq:djdT_B}, the asymmetry of $A(\omega)$ at $\omega = \mu_{L/R}$ determines \mbox{\djdTns$_B$}.

In the lower panels of \fig{fig:spectral_djdT_VG1} we show results for 
the response of the spectral function to an infinitesimal temperature difference, more specifically the integrand $\partial j(\omega) / \partial (\dT)_A$ of \eq{eq:djdT_A} which is proportional to $\partial A(\omega) / \partial (\dT)$ (evaluated numerically) times the transport window $f_R(\omega)-f_L(\omega)$. 
$\partial j(\omega) / \partial (\dT)_A$ is maximal for $\phi \sim 0.5\,\Gamma$ and vanishes in the limit $\phi\to0$ due to the narrowing transport window. The main characteristic features are a positive peak at $\omega \approx -\phi/2$ and a negative one at $\omega \approx \phi/2$, which 
appear in correspondence of the position of the leads' chemical potentials and thus of the split Kondo peaks. 
Hence the Kondo peaks strongly affect the response $\partial A(\omega) / \partial (\dT)$ to a temperature difference. 
In the limit of large $\phi$ and $U$ a simple picture emerges: the dominating effect of \DT is that the split Kondo peaks are asymmetrically enhanced or suppressed. This effect can be clearly seen for $U=8\,\Gamma$ and $\phi=4\,\Gamma$, for instance, where  $\partial j(\omega) / \partial (\dT)_A$ exhibits only two peaks at $\omega \approx \pm\phi/2$ and no response in the region in between. The sign of the two peaks is determined by the direction in which the temperature difference is applied. Since we considered $T_L = T - \dT/2$ and $T_R = T + \dT/2$, analogously to $\mu_L = - \phi/2$ and $\mu_R = + \phi/2$, it is intuitive
 that the Kondo peak at $\omega<0$ is enhanced and the one at $\omega>0$ is suppressed. For lower values of the interaction strength the behavior is more complex since the split Kondo peaks are less pronounced and merge with the Hubbard bands already for rather small values of $\phi$.
 In all cases, the overall sign when integrating $\partial j(\omega) / \partial (\dT)_A$ 
is positive (see also \fig{fig:djdT_VG1}), since for $V_G>0$ the Kondo peak at $\omega<0$ is enhanced by the proximity of the lower Hubbard band.

It is interesting to note that the nearly constant \djdT for $\phi\gtrsim\Gamma$ and $U\geq6\,\Gamma$, as observed in \fig{fig:djdT_VG1}, is dominated by the response of the split Kondo peaks. As discussed above, both contributions \mbox{\djdTns$_B$} and \mbox{\djdTns$_A$} are in this parameter regime essentially determined by the asymmetry of $A(\omega)$ and its response at $\omega = \pm\phi/2$, which corresponds to the positions of the split Kondo peaks.

In general, the Kondo resonance is suppressed by both a finite temperature $T$ and a bias voltage $\phi$,{ otherwise} the two external parameters act in a very different manner: a finite $T$ (at $\phi=0$) induces a lowering of the Kondo resonance peak height, whereas a finite $\phi$ (at $T<T_K$) {in addition} splits the peak in two, see also upper panel of \fig{fig:spectral_djdT_VG1}. From previous results it is known that the differential conductance \djdV is not affected by this difference, since it shows qualitatively the same functional dependence  on $T$ as on $\phi$~\cite{pl.sc.12,sm.gr.11,sm.gr.13,mu.bo.13,re.pl.14,mo.mo.15,an.do.16}. The question arises whether this holds true also for \djdTns. 

To address this issue, we show results for \djdT as a function of $\phi$ and $T$ in \fig{fig:djdT_U6}, for $U=6\,\Gamma$ and for various $V_G$. The nonequilibrium calculations were obtained with AMEA and the equilibrium ones with NRG~\footnoterecall{noteNRG}. Already from the equilibrium behavior displayed in the inset on the right one notices that the temperature dependence of \djdT is quite involved. Apart from $V_G=2\,\Gamma$ all curves exhibit sign changes~\footnoterecall{mixedValence}. These findings are in agreement with the results reported in \tcite{co.zl.10}, where the Seebeck coefficient $\cals$, i.e. \eq{eq:dS} for $j=0$, was investigated in detail for $\phi=0$. Note that the sign of \djdT determines the sign of $\cals$. In general, \djdT vanishes for $V_G\to0$ and also for $T\to0$. 
The latter can be understood from a Sommerfeld expansion of \eq{eq:djdT_B}~\cite{co.zl.10,ki.za.12} 
which yields 
{for \djdT a linear behavior in $T$ for  $T\to 0$. 
The vanishing \djdT for small $T$ can be observed in the inset of the right panel in \fig{fig:djdT_U6}. In contrast, the left panel reveals that for $\phi\to 0$ the thermoelectric property \djdT approaches a finite, even $V_{G}$-dependent value. This is a first indication
of the different dependence of \djdT on temperature and on bias voltage.} 

In order to investigate the $\phi$-dependence of \djdT in the Kondo regime we consider $T=0.1\,\Gamma$, which corresponds approximately to the position of the maximum of \djdT {as a function of temperature} (for $T<T_K$) 
(see inset in right panel).
The results for the dependence on $\phi$ are shown in the first two panels of \fig{fig:djdT_U6}.
One can see that the characteristic scale that governs the decrease of \djdT with $\phi$ is $\Gamma$ and not the Kondo temperature $T_K$. The appearance of additional energy scales proportional to $\Gamma$ has been also found in the equilibrium thermoelectric properties of the SIAM~\cite{co.zl.10}. 
A comparison of the functional dependence of \djdT on $\phi/\Gamma$ (for $T=0.1\,\Gamma$) and on $T/\Gamma$ (for $\phi=0$) is provided in the  second and third panel of \fig{fig:djdT_U6},{ both with logarithmic abscissa starting at $10^{-1}$}. We note that for $T=0.1\,\Gamma$ and $\phi=0$ the results obtained from AMEA and from NRG agree very well. The equilibrium ($T$-dependent) and nonequilibrium ($\phi$-dependent) data behave similarly only for large values of the gate voltage ($V_G=2\,\Gamma$), while they deviate from each other as a function of $\phi$ or $T$ for small $V_G\approx\Gamma$. In particular, in nonequilibrium the decrease of \djdT with $\phi$ is followed by a saturation to a finite (positive) value, whereas in equilibrium the oscillating behavior as a function of $T$ exhibits also sign changes. As a consequence, the resulting thermoelectric response of a SIAM in the Kondo regime is characterized by pronounced differences.

We  note that the quantity \djdT is rather sensitive to details of the spectral function $A(\omega)$. Differently to \djdVns, which involves an effective averaging over the transport window, \djdT is mainly determined by the response at $\omega = \pm \phi/2$, as discussed above. This explains the observed differences in the behavior of \djdT with respect to \djdVns, between the dependence on $T$ in equilibrium and the one of $\phi$ in nonequilibrium.

Finally, 
We stress that, while we here restrict to  linear order in
$\dT$
the present method can evaluate nonequilibrium properties for any {\em finite} $\dT$ (and $\phi$). In addition, we estimated up to what $\dT$ the linear regime 
is valid by computing a $\dT_{val}$ for which the second order term in $\dT$ 
becomes of the same order as the first order correction (we have an estimate for the second derivative of $j$ with respect to $\dT$ from finite differences). 
We get values of $\dT_{val}/T_K = {0.4, 1.5, 0.7}$ for $\phi = {0.1 \Gamma, 0.4 \Gamma, 2 \Gamma}$, for the exemplary case $U = 6 \ \Gamma$ and $V_G = \Gamma$.
Therefore, roughly for $\dT$ up to $T_K$  the linear regime is accurate.

In view of sensing applications we conclude that the response is maximal in proximity of the linear response regime for $\phi=0$ and moreover, that fairly large gate voltages are advantageous. As illustrated by the results for $V_G=2\,\Gamma$ considered here, 
a small plateau develops in the region up to $\phi\approx0.3\,\Gamma$ in which \djdT is nearly constant. In general, this might be achieved by tuning the system to a temperature just below the maximum of \djdT at $\phi=0$. 
However, further studies including also asymmetric couplings to the leads are required to fully assess the potential of quantum dot devices for nanoscale sensing applications.

\section{Conclusions}
\label{sec:conclusio}
In the present work we computed the nonequilibrium spectral and thermoelectric transport properties of a quantum dot modeled by a SIAM in presence of both, an external bias voltage $\phi$ and a temperature difference \DT between the two leads. In particular, we focused on the differential response of the current to a temperature gradient \djdT in the nonequilibrium Kondo regime ($\phi\neq0$ and $T<T_K$), for the case of an infinitesimal temperature difference \dDTns. When compared to the results for the differential conductance \djdV known from previous studies~\cite{pl.sc.12,sm.gr.11,sm.gr.13,mu.bo.13,re.pl.14,mo.mo.15,an.do.16}, we find that \djdT exhibits a more complex behavior. On the one hand, the equilibrium and nonequilibrium properties turn out to be quite different, i.e. whether one monitors \djdT as a function of $T$ or of $\phi$, and on the other hand, the relevant low-energy scales are more involved. \djdT reveals a rapid decrease with increasing $\phi$, similar to \djdVns, however, governed by a scale determined by the coupling strength $\Gamma$ and not $T_K$. This finding is consistent with equilibrium results reported in \tcite{co.zl.10}. 

A more detailed understanding is
gained by inspecting the different contributions to \djdTns, and especially by analyzing the spectral function $A(\omega)$ and its frequency-resolved response to $\Delta T$. Here, we presented
a systematic study for various values of the interaction strength $U$ and gate voltages $V_G$.
In the regime of strong interactions and sufficiently large bias voltages we provided a simple physical picture, based on the asymmetric suppression or enhancement of the split Kondo peak. A temperature reduction of \DTns/2 in one lead and the respective increase in the other one induces an asymmetric response of the two Kondo peaks.
This characteristic nonequilibrium feature lead to the qualitatively different behavior of \djdT as a function of $T$ or of $\phi$.
By extending our previous results~\cite{do.ga.15} to non-particle-hole symmetric situations we  were able to accurately resolve the asymmetric splitting of the Kondo peak in $A(\omega)$ with increasing bias voltage. We note that in contrast to other previous works (see e.g.~\tcite{du.le.13}), we here report a clear observation of this splitting, which was not accessible before, and its evolution with increasing values of $U$.

On the whole, one can conclude that \djdT is more sensitive to details in the spectral function $A(\omega)$ than the differential conductance \djdVns. Besides the fundamental interest, the findings reported here may be of value to assess the potential of quantum dots as possible nanoscale temperature sensors.

\section*{Acknowledgments}
{
We would like to thank M. Sorantin, I. Titvinidze, H. G. Evertz, T. Costi, V. Zlatic and K. Held 
for fruitful discussions. 
We acknowledge financial support from 
the Austrian Science Fund (FWF) within 
Projects F41 (SFB ViCoM) and P26508, 
 from the Simons Foundation (Many Electron Collaboration) and the Perimeter Institute for Theoretical Physics,
and from NaWi Graz.
The calculations were performed on the VSC-3 cluster Vienna as well as on the D-Cluster Graz.
Furthermore, the authors want to thank Rok {\ifmmode \check{Z}\else\v{Z}\fi{}}itko for providing his open source code NRG Ljubljana.~\footnoterecall{noteNRG}
}

\appendix
\section{Numerical differentiation of the spectral function}
\label{sec:fit}
\begin{figure*}
\begin{center}
\includegraphics[width=0.9\textwidth]{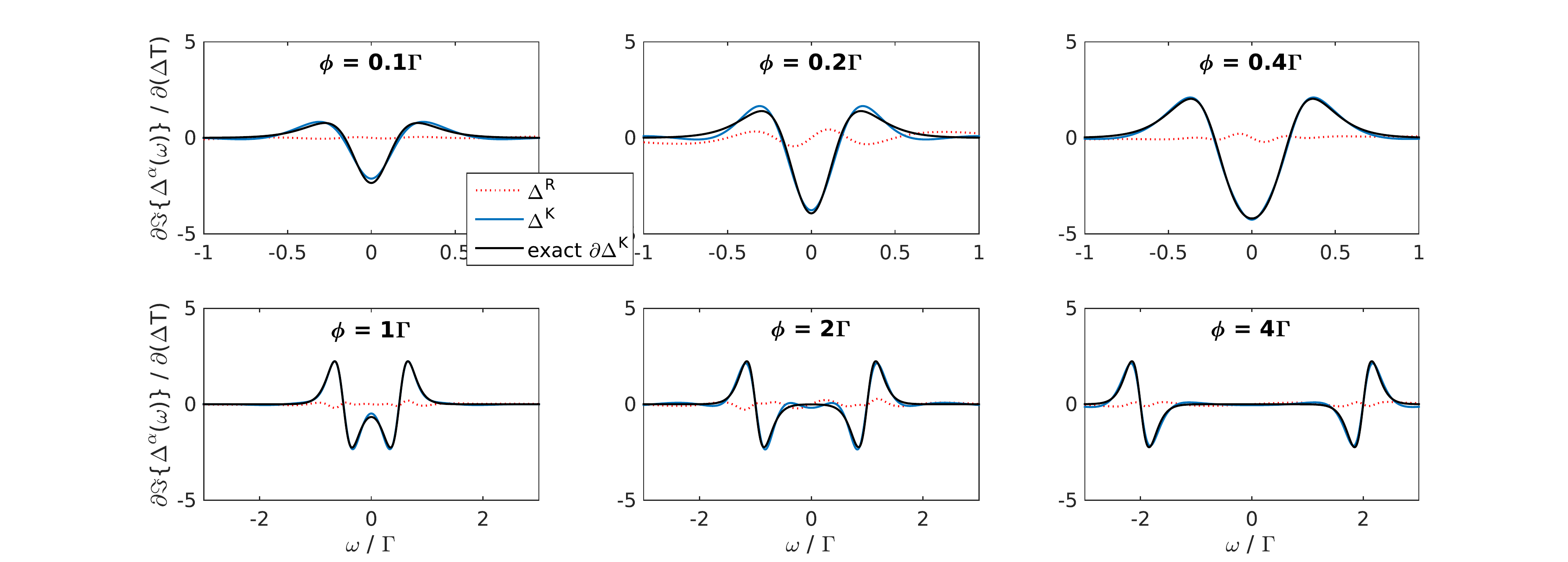}
\caption{(Color online) Numerically estimated differentials $\partial \und \Delta_\mathrm{aux} (\omega) / \partial (\dT)$, labeled by 
$\Delta^{\alpha} (\omega)$ with $\alpha=R,K$, together with the analytical expressions   
{given in} \eqs{eq:FDT} and (\ref{eq:Dhyb}). Only the exact $\partial \Delta^{K}(\omega) / \partial (\dT)$ is plotted since $\partial \Delta^{R}(\omega) / \partial (\dT) = 0$. Especially for $\phi=0.2\,\Gamma$ larger errors occur and $\partial \Delta^{R}(\omega) / \partial (\dT)$ is clearly nonzero.
}
\label{fig:fits}
\end{center}
\end{figure*}
In order to evaluate \eq{eq:djdT} in nonequilibrium $\phi\neq0$ one needs to calculate $\partial \mathcal T(\omega)/\partial(\dT)$ in \eq{eq:djdT_A} numerically. Details on this are given here. 

AMEA consists of two major steps: 
(i) the mapping procedure, and (ii) the many-body solution. In (i) we optimize all available bath parameters in the auxiliary system for a given number of bath sites  $N_B$ in order to reproduce the hybridization function of the  physical system $\und \Delta(\omega)$, \eq{eq:Dhyb}, by the auxiliary one $\und \Delta_\mathrm{aux}(\omega)$ as accurately as possible. This is done through a fit on the real $\omega$-axis and with a parallel tempering (PT) routine~\cite{ea.de.05,hu.ne.96}, see \tcite{do.ga.15}. We thus aim at solving a global optimization problem in the fit. Once the optimal bath parameters are obtained we thereafter proceed to step (ii) and treat the many-body problem. In \tcite{do.ga.15} we presented a solution strategy based on matrix product states, which we also make use of in the present work.

{$f'(\Delta T)\big|_{\Delta T=0}:=\partial \mathcal T(\omega,\Delta T)/\partial(\dT)\big|_{\Delta T=0}$} is estimated by finite differences. First we compute for a certain $\phi$ the fit for the particle-hole symmetric case at $\dT=0$, in the manner just described above (see also \tcite{do.ga.15}). Thereafter, we introduce a small finite temperature difference $\dT\approx T/10$, start the fit from the bath parameters obtained for $\dT=0$ and optimize only locally~\footnote{Such a local optimization may be done for instance within the PT scheme with reduced artificial temperatures.}. 
By this, the fit for $\dT\neq0$ is close in parameter space to the one for $\dT=0$ and we can thus estimate the linear response with respect to \DTns. In order to approximate a derivative by finite differences accurately it would be appropriate, in principle, to consider various stencils with varying values of \DTns. Since this would require a major computational effort in the many-body calculation we consider here for simplicity only two values $\dT=\pm T/10$. Note further that these two fits are connected by a particle-hole transformation. The derivative {$f'(\Delta T)\big|_{\Delta T=0}$} is approximated by the central finite difference, i.e. based on $\dT=\pm T/10$. Together with the computation for $\dT=0$ this amounts to three many-body calculations for a certain $\phi$ and $T$.

Already on the level of the mapping procedure, i.e. before solving the many-body problem, one can check how well the finite difference approach works with the chosen $N_B$ and \DTns. For this we compute the derivative of the hybridization function. The derivative of the physical hybridization function \eq{eq:Dhyb} of the original impurity problem can be carried out analytically. The temperature difference enters only through the Keldysh component, so that $\partial \Delta^R(\omega)/\partial(\dT) = 0$ and the expression for $\partial \Delta^K(\omega)/\partial(\dT)$ is obtained by differentiating the Fermi functions in \eqs{eq:Dhyb} and (\ref{eq:FDT}). A comparison of the estimated derivatives versus the exact ones is shown in \fig{fig:fits} for a representative set of bias voltages. Overall, a good agreement is found. Larger differences occur for instance at $\phi=0.2\,\Gamma$, where the retarded part is clearly nonzero and also the Keldysh part exhibits more deviations. Judging from such an analysis and from comparing the values of the forward, backward and central difference approximations of {$f'(\Delta T)\big|_{\Delta T=0}$}, we can state that the
presented data points for the numerical derivatives in \fig{fig:djdT_VG1} and \fig{fig:djdT_U6} are less reliable  for $\phi=0.2\,\Gamma$. For larger values of $\phi$ and for $\phi\leq 0.1\,\Gamma$ results are more accurate. A rigorous quantitative estimate of the error originating from the numerical derivative would require a more expensive numerical analysis.
On the contrary, the computed spectral functions in \fig{fig:spectral_djdT_VG1} can be regarded to have a high accuracy.~\footnote{In \tcite{do.ga.15} we achieved converged and accurate results with $N_B=14$ for $T=0.05\,\Gamma$. Here, we used $N_B=14$ for an increased temperature of $T=0.1\,\Gamma$, which improves the costfunction values $\mathcal{C}/\mathcal{C}_0$ by nearly one order of magnitude.} This shows that the computation of derivatives magnifies small inaccuracies of the mapping procedure, which are present for a certain finite value of $N_B$.

\end{document}